%% file: main.tex
\documentclass{article}
\usepackage{spconf,amsmath,graphicx,hyperref}
\usepackage{booktabs,multirow}
\usepackage{xcolor}
\usepackage{soul}
\sethlcolor{yellow!25}

\title{ReplayFactor: Replay-Aware Dual-Attribute Learning for Practical Speech Deepfake Detection}
\name{%
  \shortstack{%
    Miao He$^{1,3}$, Jiayi Zhou$^{3}$, Yu Zheng$^{3}$,
    Zhongjie Ba$^{1,2}$, Peng Cheng$^{1,2,\dagger}$, \\
    Yuxi Zhou$^{2}$, Junjie Zhu$^{3}$,
    Changhao Zhang$^{3}$, Jian Liu$^{3,\dagger}$
  }%
}

\address{
  $^{1}$ The State Key Laboratory of Blockchain and Data Security, Zhejiang University  \\
  $^{2}$ Hangzhou High-Tech Zone (Binjiang) Institute of Blockchain and Data Security \\
  $^{3}$ Ant Group
}
\begin{document}
\ninept
\maketitle

\begin{abstract}
Advances in synthesis techniques have made deepfake speech increasingly realistic, amplifying voice security threats, including unauthorized voice cloning and impersonation. Although existing deepfake detectors perform well on digitally generated speech, their performance often degrades on physically replayed deepfakes, which constitute a low-cost and practical threat to real-world systems. Existing research focuses primarily on dataset construction and evaluation, while available dataset diversity and replay-aware detector design remain limited. We propose ReplayFactor, a dual-attribute framework that enhances replay evidence by fusing self-supervised learning (SSL) representations with an acoustic prior, while leveraging multi-task learning and supervised contrastive learning to promote robust, generalizable representations. We also construct ReplayFake-CN, a large-scale, multi-device Chinese dataset comprising approximately 185k replayed synthetic utterances to complement public benchmarks. Our framework improves macro-F1 from 60.99\% to 76.44\% and reduces EER from 11.86\% to 3.05\% on the EchoFake open set over the baseline, demonstrating strong potential for practical replayed deepfake detection.
\end{abstract}

\begin{keywords}
speech deepfake detection, physical replay,
acoustic prior,  supervised contrastive learning
\end{keywords}

\section{Introduction}
\label{sec:intro}

Advances in speech synthesis have enabled highly realistic speech generation, raising concerns about the reliability of voice-based services and voice impersonation. As a countermeasure, self-supervised learning (SSL) based speech deepfake detectors have achieved strong performance on digitally generated speech~\cite{tak2022automatic,ge2025post,liu2025nes2net,dowerah2026speechBoard}, but their performance often degrades when synthetic speech is played through a loudspeaker and re-recorded by a microphone~\cite{delgado2026deepfake}. Such physical replay deepfake attacks constitute a low-cost and practical threat to Automatic Speaker Verification systems in real-word, yet have received limited research attention.

Physical replay introduces various distortions through the playback device, acoustic environment, and recording conditions, which can alter or mask synthesis artifacts~\cite{muller2025replaydf}. These distortions can shift feature distributions and cause existing detectors to misclassify replayed synthetic speech as genuine~\cite{delgado2026deepfake,muller2025replaydf}.
Moreover, Variations in synthesis models and playback--recording configurations further challenge detection generalization to unseen conditions~\cite{zhang2025echofake}.


Recent works ReplayDF~\cite{muller2025replaydf} and EchoFake~\cite{zhang2025echofake} provide public benchmarks for replayed synthetic speech, but their coverage of synthesis methods, replay devices, and languages remains limited. Evaluations on these benchmarks show that detectors trained on non-replayed speech can suffer substantial performance degradation under replay. Even when trained on replayed synthetic speech, detectors still struggle to generalize to unseen synthesis methods and replay devices~\cite{zhang2025echofake}. Existing approaches primarily rely on data augmentation~\cite{delgado2026deepfake} or replay simulation~\cite{sanchez2026rirplay} to improve robustness to replayed speech, while replay-aware detector design  remains underexplored, especially for cross-condition generalization.


To address these limitations, we propose ReplayFactor, a dual-attribute framework for replay-aware speech deepfake detection. At the feature level, we fuse the SSL-based representation with an acoustic prior to  enhance replay cues. For representation learning, it combines multi-task learning with dual-space supervised contrastive learning to promote the cross-condition consistency of source and replay attributes and improve generalization. To complement existing public benchmarks, We also construct ReplayFake-CN\textsuperscript{*}, a large-scale Chinese dataset comprising approximately 185k replayed synthetic utterances across 90 playback--recording device pairs. Experiments on EchoFake, ReplayDF, and ReplayFake-CN validate the effectiveness and generalization of the proposed framework. On the EchoFake open set, compared with the baseline reported in EchoFake~\cite{zhang2025echofake}, our method improves macro-F1 from 60.99\% to 76.44\% and reduces EER from 11.86\% to 3.05\%, demonstrating strong potential for realistic replayed deepfake detection.

{\renewcommand{\thefootnote}{\fnsymbol{footnote}}
\footnotetext[1]{The full dataset will be released soon at
\url{https://huggingface.co/datasets/Ahola/ReplayFake-CN}.}
}
\begin{figure*}[!t]
  \centering
  \includegraphics[width=\textwidth]{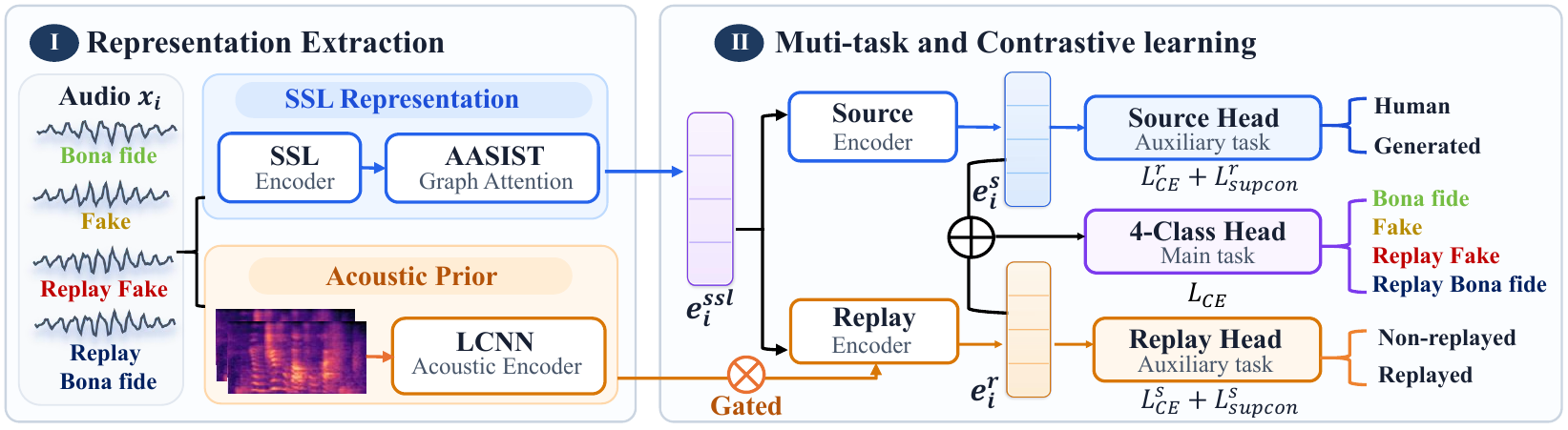}
  \caption{Overview of the proposed framework, comprising (I) Representation extraction with SSL and acoustic-prior fusion; and (II) multi-task and supervised contrastive learning.}
  \label{fig:framework}
  \vspace{-0.3cm}
\end{figure*}

\section{Related Work}
\label{sec:related}

\noindent\textbf{Speech Deepfake Detection.}
Speech deepfake detection has evolved from handcrafted acoustic features utilization~\cite{sahidullah2015comparison,todisco2016cqcc,delgado2018asvspoof}, to end-to-end models such as RawNet2~\cite{tak2021end} and AASIST~\cite{jung2022aasist}, and more recently to SSL-based detectors with improved transferability~\cite{tak2022automatic,ge2025post,liu2025nes2net}.
However, most studies focus on non-replay synthetic speech, with limited attention to realistic scenarios where synthetic speech is physically replayed~\cite {delgado2026deepfake}. 

\noindent\textbf{Replay-Aware Deepfake Datasets and Detection.} ReplayDF~\cite{muller2025replaydf} and EchoFake~\cite{zhang2025echofake} provide public benchmarks for replayed synthetic speech. However, ReplayDF covers only a limited set of synthesis methods, while EchoFake is English-only and includes few replay devices. Neither benchmark covers Chinese speech or replay through smart wearable devices such as smart glasses. Existing studies primarily analyze replay-induced degradation~\cite{delgado2026deepfake} or simulate replay effects~\cite{sanchez2026rirplay}, while replay-aware detector design remains underexplored. Even when detectors are trained on replayed speech, generalization to unseen synthesis methods and replay devices remains challenging~\cite{zhang2025echofake}. To address these gaps, we introduce ReplayFake-CN, a large-scale Chinese dataset of replayed synthetic speech to complement existing datasets, and propose a dual-attribute learning framework for replay-aware deepfake detection.

\input{proposed_method}

\section{ReplayFake-CN Dataset}
\label{sec:dataset}

To complement existing public benchmarks~\cite{muller2025replaydf,zhang2025echofake} and systematically evaluate replay-aware deepfake detection across more synthesis methods and replay devices, we introduce ReplayFake-CN, a large-scale multi-device Chinese dataset containing approximately 185k replayed synthetic utterances (232.5 hours). 
The source corpus comprises recordings from 1,921 speakers, while the texts cover daily-life and payment-related content. We employ 22 state-of-art TTS models to generate deepfake speech, such as CosyVoice3~\cite{du2025cosyvoice} and LLaSA-1B~\cite{ye2025llasa}, etc. To simulate practical spoofing scenarios, the generated speech is physically played through 9 loudspeakers and re-recorded using 8 phones and 2 smart glasses, yielding 90 playback--recording device pairs. More details are provided in Table~\ref{tab:replayfake_stats}.
Each utterance is released with fine-grained metadata, including speaker ID, TTS model, playback and recording devices, distance, and text. 
We also provide training, development, and evaluation protocols, including out-of-domain protocols with unseen synthesis models, unseen devices, and unseen model--device combinations, enabling evaluation of cross-domain generalization.

\begin{table}[t]
\centering
\caption{Statistics of ReplayFake-CN.}
\label{tab:replayfake_stats}
\small
\setlength{\tabcolsep}{1.5pt}
\renewcommand{\arraystretch}{1.05}
\begin{tabular*}{\columnwidth}
{@{\extracolsep{\fill}}p{0.25\columnwidth}p{0.72\columnwidth}@{}}
\toprule
\textbf{Variable} & \textbf{Configuration} \\
\midrule
Data scale
& 184,954 utterances, 232.5 hours \\

Source speech
& Utterances from 1,921 speakers \\

Language
& Chinese \\

Text
& Daily-life and payment-related scenarios \\

TTS models
& 22 models, e.g., CosyVoice3~\cite{du2025cosyvoice}, F5-TTS~\cite{chen2025f5}, Step-Audio~\cite{huang2025step}, LLaSA-1B~\cite{ye2025llasa}, and  etc. \\

Play devices
& 6 consumer and 3 high-fidelity loudspeakers, e.g., JBL, HP-speaker, Edifier-MR4, etc. \\

Record devices
& 8 phones and 2 smart-glasses, e.g., iPhone,  OPPO, HuaweiMate, and Rokid\_glasses, etc. \\

Combinations
& 90 playback--recording device pairs \\

Distance
& 10, 30, and 50\,cm \\
\bottomrule
\end{tabular*}
\vspace{-0.4cm}
\end{table}

\input{experiment}

\section{Conclusion}
In this paper, we propose a dual-attribute learning framework for replay-aware speech deepfake detection that explicitly learns source and replay attributes by enhancing SSL representations
with an acoustic prior and combining multi-task learning with supervised contrastive learning.
We also construct ReplayFake-CN, a large-scale Chinese dataset covering diverse synthesis methods and replay devices, to broaden existing datasets and support our evaluation of cross-condition generalization, facilitating future research.
Experiments on two public datasets and ReplayFake-CN validate the effectiveness of our framework, highlighting its potential for replayed deepfake attack detection in realistic scenarios.

\bibliographystyle{IEEEbib}
\bibliography{strings,refs}

\end{document}

%% file: proposed_method.tex

\section{Proposed Method}
\label{sec:method}

\subsection{Problem Formulation}
\label{sec:problem}
As illustrated in Fig.~\ref{fig:framework}, each utterance $x_i$ is assigned with a label $y_i\in\{B,F, RB, RF\}$, denoting bona fide, fake, replayed bona fide, and replayed fake speech, respectively. We factorize $y_i$ into a source attribute $y_i^s$, distinguishing human speech $\{B,RB\}$ from generated speech $\{F,RF\}$, and a replay attribute $y_i^r$, distinguishing non-replayed speech $\{B,F\}$ from replayed speech $\{RB,RF\}$. This factorization enables explicit learning of each attribute across variations in the other.
The framework consists of two components: (1) \emph{Replay-Aware Feature Fusion}, which enhances SSL representations with an acoustic prior through gated fusion; and (2) \emph{Multi-Task and Contrastive Learning}, which jointly optimizes the four-class main classification task and two auxiliary attribute tasks, with dual-space contrastive learning to promote source and replay attribute consistency across diverse conditions.

\subsection{Replay-Aware Feature Fusion}
\label{sec:fusion}

We first extract a shared representation $e_i^{ssl}$ using a pretrained SSL encoder followed by an AASIST graph backend. Then, two attribute-specific encoders map this representation into a source embedding $e_i^s$ and a replay embedding $e_{i,\mathrm{ssl}}^r$, respectively. The source branch targets human-versus-generated discrimination, whereas the replay branch targets non-replayed-versus-replayed discrimination. To strengthen the replay branch, we further introduce an acoustic prior that complements SSL representations with low-level replay-sensitive cues, such as subtle high-frequency differences and channel response distortions~\cite{witkowski2017highfreq,avila2019channel}.



Specifically, we used constant-Q cepstral coefficients (CQCCs) as the acoustic prior, which have been shown to be effective in replay attack detection~\cite{todisco2016cqcc,delgado2018asvspoof}. We concatenate the static  coefficient $c_i$ with their first- and second-order temporal derivatives and encode them using an LCNN~\cite{kotta2020subband}. They are computed as:
\begin{equation}
\begin{split}
c_i&=\operatorname{DCT}\!\left(
\operatorname{Resample}
\left(\log\left(|\operatorname{CQT}(x_i)|^2+\epsilon\right)\right)
\right),\\
q_i&=[c_i,\Delta c_i,\Delta^2c_i],\quad
e_i^a=\operatorname{LCNN}(q_i).
\end{split}
\label{eq:cqcc}
\end{equation}
Here, $\epsilon$ is a small constant for numerical stability, and $e_i^a$ provides the acoustic representation used to enhance the replay branch.

We fuse the acoustic embedding into the SSL-based replay embedding through a sample-wise gate:
\begin{equation}
e_i^r=e_{i,\mathrm{ssl}}^r+g_i e_i^a,
\end{equation}
where $g_i\in(0,1)$ adaptively controls the acoustic contribution for each utterance. This residual formulation retains the SSL-based representation while incorporating complementary acoustic cues. The resulting replay embedding $e_i^r$, together with the source embedding $e_i^s$, is used for the classification and contrastive objectives below.


\subsection{Multi-Task and Dual-Space Contrastive Learning}
\label{sec:objective}



We jointly optimize a four-class classification task and two auxiliary attribute classification tasks. The main classifier takes the concatenated representation $[e_i^s;e_i^r]$ to distinguish $B$, $F$, $RB$, and $RF$. The source and replay classifiers take $e_i^s$ and $e_i^r$, respectively, to predict the corresponding binary attribute labels $y_i^s$ and $y_i^r$. Their cross-entropy losses are denoted by $\mathcal{L}_{\mathrm{CE}}^4$, $\mathcal{L}_{\mathrm{CE}}^s$, and $\mathcal{L}_{\mathrm{CE}}^r$.

Beyond classification supervision, we further apply supervised contrastive learning~\cite{khosla2020supervised} separately in the source and replay spaces to encourage consistency within each attribute. In the source space, $B$ and $RB$ share the human label, while $F$ and $RF$ share the generated label, encouraging source consistency across  diverse replay states. In the replay space, $B$ and $F$ share the non-replayed label, while $RB$ and $RF$ share the replayed label, encouraging replay consistency across source types.

For a mini-batch $\mathcal{B}$ of size $N$, the positive set for anchor $i$ in space $u\in{s,r}$ is
$P_i^u=\{p\in\mathcal{B}\setminus{i}:y_p^u=y_i^u\}$.
Let $z_i^u$ denote the $\ell_2$-normalized contrastive representation in space $u$, and define
$s_{ij}^u=(z_i^u)^\top z_j^u/\tau_u$ as the temperature-scaled cosine similarity, where $\tau_u$ is the temperature. The supervised contrastive loss is:
\begin{equation}
\mathcal{L}_{\mathrm{sup}}^u
-\frac{1}{N}\sum_{i=1}^{N}
\frac{1}{|P_i^u|}
\sum_{p\in P_i^u}
\log
\frac{\exp(s_{ip}^u)}
{\sum_{a\in\mathcal{B}\setminus{i}}\exp(s_{ia}^u)}.
\end{equation}

The overall training objective is:
\begin{equation}
\mathcal{L}
=\mathcal{L}_{\mathrm{CE}}^4
+\alpha_s\mathcal{L}_{\mathrm{CE}}^s
+\alpha_r\mathcal{L}_{\mathrm{CE}}^r
+\lambda_s\mathcal{L}_{\mathrm{sup}}^s
+\lambda_r\mathcal{L}_{\mathrm{sup}}^r.
\label{eq:total}
\end{equation}
where $\alpha_s,\alpha_r$ weight the auxiliary classification losses and $\lambda_s,\lambda_r$ weight the contrastive losses. Together, these objectives supervise four-class discrimination while encouraging attribute-specific consistency in the two embedding spaces.

%% file: experiment.tex
\begin{table}[t]
\centering
\caption{Results (\%) on EchoFake and ReplayDF. Closed and Open denote the EchoFake evaluation sets. ReplayDF reports $RB$ and $RF$ F1 scores from EchoFake-trained models.}
\label{tab:public_results}
\small
\setlength{\tabcolsep}{1.5pt}
\renewcommand{\arraystretch}{1.05}
\begin{tabular*}{\columnwidth}
{@{\extracolsep{\fill}}llcccccc@{}}
\toprule
\multirow{2}{*}{Model} &
\multirow{2}{*}{Test} &
\multirow{2}{*}{EER$\downarrow$} &
\multicolumn{5}{c}{F1-score$\uparrow$} \\
\cmidrule(lr){4-8}
& & & macro. & $B$ & $RB$ & $F$ & $RF$ \\
\midrule
\multirow{3}{*}{\shortstack[c]{XLSR-Linear\\ $^\dagger$Reported in~\cite{zhang2025echofake}}}
& Closed   & 0.27  & \textbf{98.81} & 99.63 & \textbf{98.16}
& 99.43 & \textbf{98.02} \\
& Open     & 11.86 & 60.99 & 78.83 & 40.13 & 75.64 & 49.34 \\
& ReplayDF & -- & -- & -- & 7.72 & -- & 19.02 \\
\midrule
\multirow{3}{*}{XLSR-AASIST}
& Closed   & 0.47 & 98.33 & 98.68 & 97.97 & 98.66
& \textbf{98.02} \\
& Open     & 8.17 & 56.99 & 86.69 & 19.96 & 71.91 & 49.40 \\
& ReplayDF & -- & -- & -- & 7.02 & -- & 13.69 \\
\midrule
\multirow{3}{*}{WavLM-Nes2Net}
& Closed & 1.68 & 76.75 & 78.15 & 66.01 & 84.08 & 78.77 \\
& Open & 21.72 & 46.13 & 67.84 & 20.45 & 56.81 & 39.42 \\
& ReplayDF & -- & -- & -- & 0.82 & -- & 19.75 \\
\midrule
\multirow{3}{*}{MMS-Linear}
& Closed   & \textbf{0.13} & 97.37 & \textbf{99.73} & 94.81
& \textbf{99.70} & 95.24 \\
& Open     & 8.09 & 59.80 & 88.09 & 8.79 & \textbf{82.99} & 59.34 \\
& ReplayDF & -- & -- & -- & 1.97 & -- & 18.26 \\
\midrule
\multirow{3}{*}{MMS-AASIST}
& Closed   & 0.54 & 98.40 & 99.30 & 97.59 & 99.17 & 97.53 \\
& Open     & 7.26 & 62.27 & 90.14 & 25.93 & 77.99 & 55.02 \\
& ReplayDF & -- & -- & -- & 6.73 & -- & 25.94 \\
\midrule
\multirow{3}{*}{Ours}
& Closed   & 0.53 & 97.93 & 99.04 & 97.16 & 98.72 & 96.81 \\
& Open     & \textbf{3.05} & \textbf{76.44} & \textbf{93.91}
& \textbf{61.12} & 82.01 & \textbf{68.72} \\
& ReplayDF & -- & -- & -- & \textbf{38.31} & --
& \textbf{40.32} \\
\bottomrule
\end{tabular*}
\vspace{-0.4cm}
\end{table}

\begin{table}[!t]
\centering
\caption{Accuracy (\%) on ReplayFake-CN under unseen conditions.
Binary and four-class models are reported separately.}
\label{tab:replayfake_results}
\small
\setlength{\tabcolsep}{2.0pt}
\renewcommand{\arraystretch}{1.05}
\begin{tabular*}{\columnwidth}
{@{\extracolsep{\fill}}lcccc@{}}
\toprule
Method &
\shortstack{Overall\\ACC} &
\shortstack{Unseen\\Synthesis} &
\shortstack{Unseen\\Devices} &
\shortstack{Unseen\\Synth.--Devices} \\
\midrule
\multicolumn{5}{@{}l}{%
\emph{Binary public checkpoint: AntiDeepfake~\cite{ge2025post}}} \\
MMS-300M
& 55.43 & 64.04 & 46.63 & 62.98 \\
\midrule

MMS-AASIST
& 84.22 & 93.97 & 82.59 & 73.30 \\
Ours
& \textbf{85.44} & \textbf{96.95}
& \textbf{83.00} & \textbf{73.75} \\
\bottomrule
\end{tabular*}
\vspace{-0.4cm}
\end{table}

\section{Experiments}
\label{sec:experiments}

\subsection{Experimental Setup}
\label{sec:setup}

\noindent\textbf{Datasets and Evaluation.}
We evaluate on EchoFake~\cite{zhang2025echofake}, ReplayDF~\cite{muller2025replaydf}, and ReplayFake-CN. For EchoFake, we follow the official Train, Dev, Eval-closed, and Eval-open splits to assess performance under both closed and open conditions. For cross-dataset evaluation, models trained on EchoFake are directly tested on ReplayDF, which contains only $RB$ and $RF$. For ReplayFake-CN, we train models using its $RF$ training split together with separate, non-public internal data from the $B$, $F$, and $RB$ classes, nearly 400k samples in total. We report $RF$ classification accuracy on the ReplayFake-CN eval split under unseen synthesis models, unseen devices, and unseen model–device combinations, with the held-out conditions excluded from all training data.
We report F1-score for four-class classification and binary EER for anti-spoofing, treating $B$ as bona fide and ${F,RB,RF}$ as spoofing attacks.

\noindent\textbf{Baselines.}
We compare with XLSR-Linear (baseline in~\cite{zhang2025echofake}), XLSR-AASIST~\cite{tak2022automatic}, AntiDeepfake~\cite{ge2025post}, and Nes2Net~\cite{liu2025nes2net}. For a fair comparison, we replace their binary output layer with four-class and reproduce all methods under the same protocol and settings.

\noindent\textbf{Implementation Details.}
We use MMS-300M from AntiDeepfake~\cite{ge2025post} as SSL encoder and fully fine-tune it with an AASIST backend using Adam with a weight decay of $10^{-4}$. The learning rate is $10^{-6}$ for SSL and AASIST, and $10^{-4}$ for the LCNN. Each mini-batch contains all four classes, and the checkpoint with the highest development macro-F1 is selected for evaluation.

\subsection{Main Results}
\label{sec:main_results}

\noindent\textbf{Performance on public datasets.}
Table~\ref{tab:public_results} summarizes the results on EchoFake and ReplayDF. Although our method yields slightly lower closed-set performance, it achieves substantial gains on the open set. Compared with XLSR-Linear as reported in EchoFake~\cite{zhang2025echofake}, our method reduces EER from 11.86\% to 3.05\% and improves macro-F1 from 60.99\% to 76.44\%, demonstrating stronger generalization to unseen conditions.
On ReplayDF, our method achieves F1 scores of 38.31\% and 40.32\% for $RB$ and $RF$, respectively, outperforming the corresponding best baseline scores of 6.73\% and 25.94\%. Nevertheless, the cross-dataset performance remains limited, indicating that generalization across languages, synthesis methods, and playback devices remains challenging.

\noindent\textbf{Performance on ReplayFake-CN.}
Table~\ref{tab:replayfake_results} reports results on ReplayFake-CN under unseen synthesis models, unseen devices, and jointly unseen synthesis models and devices. The public AntiDeepfake~\cite{ge2025post} checkpoint, evaluated without fine-tuning, achieves 55.43\% low binary detection accuracy, indicating limited detection effectiveness on ReplayFake-CN. MMS-AASIST and our method are trained on the same large training set, comprising the $RF$ from ReplayFake-CN training split and separate internal $B$, $F$, and $RB$ data. Our method consistently outperforms MMS-AASIST across all three protocols, improving overall $RF$ classification accuracy from 84.22\% to 85.44\%. The largest gain is observed under unseen synthesis models, increasing from 93.97\% to 96.95\%, while accuracy reaches 83.00\% and 73.75\% under unseen devices and jointly unseen conditions, indicating that device-related domain shifts remain challenging.
The modest overall gain may partly reflect the substantially larger training set (400k vs 40k samples on EchoFake), which could strengthen the baseline and leave less room for additional gains from model design. Our method nevertheless provides consistent improvements in this setting and achieves larger gains in the smaller-training-set EchoFake experiments (Table~\ref{tab:public_results}).

\subsection{Ablation Study}
\label{sec:ablation}

Table~\ref{tab:ablation} conducts an ablation study on the EchoFake open set. We first compare our model with a four-class baseline using the same MMS-AASIST backbone, and then remove individual components from the full model. The \emph{w/o acoustic prior} variant removes the acoustic prior and gated fusion, while \emph{w/o SupCon} removes the dual-space contrastive learning. Finally, \emph{w/o 4-class head} retains the two attribute heads but removes the 4-class head.

\begin{table}[t]
\centering
\caption{Ablation results on the EchoFake open set (\%).}
\label{tab:ablation}
\small
\setlength{\tabcolsep}{1.5pt}
\renewcommand{\arraystretch}{1.05}
\begin{tabular*}{\columnwidth}
{@{\extracolsep{\fill}}lcccccc@{}}
\toprule
\multirow{2}{*}{Configuration}
& \multirow{2}{*}{EER$\downarrow$}
& \multicolumn{5}{c}{F1-score$\uparrow$} \\
\cmidrule(lr){3-7}
& & macro-F1 & $B$ & $RB$ & $F$ & $RF$ \\
\midrule
4-class baseline
& 7.26 & 62.27 & 90.14 & 25.93 & 77.99 & 55.02 \\

\textbf{Ours (full)}
& 3.05 & 76.44 & 93.91 & 61.12 & 82.01 & 68.72 \\
\midrule

w/o acoustic prior
& 8.11 & 73.54 & 89.07 & 61.44 & 84.28 & 59.37 \\

w/o SupCon
& 2.70 & 73.89 & 93.82 & 50.77 & 82.85 & 68.14 \\

w/o 4-class head
& 4.31 & 69.22 & 94.84 & 32.97 & 84.03 & 65.03 \\
\bottomrule
\end{tabular*}
\vspace{-0.4cm}
\end{table}

The full model reduces EER from 7.26\% to 3.05\% and improves
macro-F1 from 62.27\% to 76.44\% over the 4-class baseline, demonstrating the overall benefit of dual-attribute auxiliary learning.
Removing the acoustic prior increases EER to 8.11\% with obvious drop of F1 score on $RF$, suggesting that low-level acoustic prior contributes replay-sensitive evidence beyond SSL features. 
Removing SupCon slightly lowers EER but reduces macro-F1 to 73.89\%, mainly due to the drop in $RB$ F1, suggesting that it helps maintain source--replay attribute consistency for this challenging class.
Finally, removing the 4-class head causes the largest macro-F1 drop, suggesting that joint supervision task complements the two attribute tasks by capturing source--replay interactions.

\subsection{Further Analysis}
\label{sec:further_analysis}
\begin{figure}[t]
    \centering
    \includegraphics[width=\columnwidth]{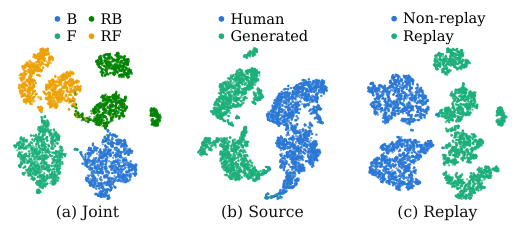}
    \caption{t-SNE visualizations on the EchoFake closed set:
    (a) joint, (b) source, and (c) replay representations.}
    \label{fig:tsne}
    \vspace{-0.4cm}
\end{figure}

\noindent\textbf{Representation Visualization.}
Figure~\ref{fig:tsne} presents t-SNE visualizations on the EchoFake closed set.
The representations exhibit clear overall separation among the
joint four classes, as well as source and replay attribute spaces. Nevertheless, slight overlap remains between $RB$ and $RF$ in the joint space, suggesting that distinguishing human from synthetic speech under physical replay remains challenging.

\begin{figure}[t]
    \centering
    \includegraphics[width=\columnwidth]{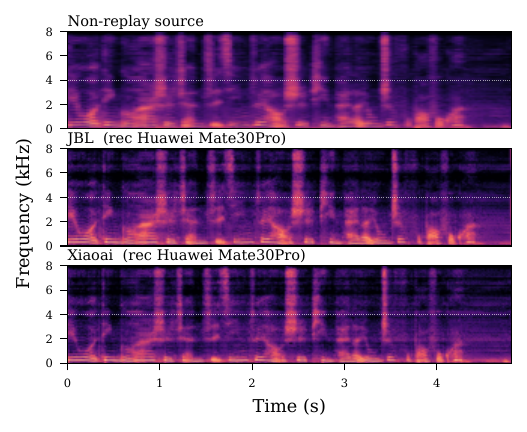}
    \vspace{-0.7cm}
    \caption{Spectrograms of the original non-replay utterance (top) and its replays through JBL (middle) and Xiaoai (bottom) loudspeakers, both recorded with a Huawei Mate30Pro phone.}
    \label{fig:stft}
    \vspace{-0.4cm}
\end{figure}
\noindent\textbf{Replay Effect Analysis.}
We also analyze replay effects across playback devices, recording devices, and TTS systems. 
As shown in Fig.~\ref{fig:stft}, the same utterance played by the JBL and Xiaoai loudspeaker exhibits distinct patterns of high-frequency attenuation, particularly above 4~kHz, suggesting that replay affects high-frequency energy differently across loudspeakers.
Beyond the illustrated examples, we further observe that recording devices also introduce different spectral distortions, with smart glasses tending to exhibit greater frequency-response distortion than smartphones.
Moreover, synthesis artifacts from different TTS models are affected by replay to varying degrees.
These observations highlight the need for future research to jointly consider device and synthesis characteristics when developing robust replay-aware deepfake detectors.
